\documentclass[article,sort&compress]{elsarticle}

\usepackage{hyperref}

\journal{Applied Energy}

\begin{document}

\begin{frontmatter}

\title{Open Power System Data - Frictionless data for electricity system modelling}

\author[dtu]{Frauke Wiese}

\author[basel,neon]{Ingmar Schlecht\corref{cor1}}
\ead{ingmar.schlecht@unibas.ch}
\cortext[cor1]{Corresponding author}

\author[euf]{Wolf-Dieter Bunke}

\author[tub]{Clemens Gerbaulet}

\author[neon,hsog,mercator]{Lion Hirth}

\author[euf]{Martin Jahn}

\author[diw,tennet]{Friedrich Kunz}

\author[tub,diw,sru]{Casimir Lorenz}

\author[neon,hu]{Jonathan M\"uhlenpfordt}

\author[neon,tub]{Juliane Reimann}

\author[diw]{Wolf-Peter Schill}

\address[dtu]{Technical University of Denmark, DTU Management Engineering, Produktionstorvet, Building 426, 2800 Kongens Lyngby, Denmark}
\address[basel]{University of Basel, Faculty of Business and Economics, Peter Merian-Weg 6, 4052 Basel, Switzerland}
\address[neon]{Neon Neue Energie\"okonomik GmbH (Neon), Karl-Marx-Platz 12, 12043 Berlin, Germany}
\address[euf]{Europa-Universit\"at Flensburg, Energy and Environmental Management, Auf dem Campus 1, 24943 Flensburg, Germany}
\address[tub]{Technische Universit\"at Berlin, Stra\ss{}e des 17. Juni 135, 10236 Berlin, Germany}
\address[hsog]{Hertie School of Governance, Friedrichstra{\ss}e 180, 10117 Berlin, Germany}
\address[mercator]{Mercator Research Institute on Global Commons and Climate Change, Torgauer Stra{\ss}e 12, 10829 Berlin}
\address[diw]{German Institute for Economic Research (DIW Berlin), Mohrenstra{\ss}e 58, 10117 Berlin, Germany}
\address[tennet]{TenneT TSO GmbH, Bernecker Stra{\ss}e 70, 95448 Bayreuth}
\address[sru]{Sachverst\"andigen Rat f\"ur Umweltfragen, Luisenstra\ss{}e 46, 10117 Berlin, Germany}
\address[hu]{Humboldt-Universit\"at zu Berlin, School of Business and Economics, Spandauer Stra{\ss}e 1, 10178 Berlin, Germany}

\begin{abstract}
    The quality of electricity system modelling heavily depends on the input data used. Although a lot of data is publicly available, it is often dispersed, tedious to process and partly contains errors. We argue that a central provision of input data for modelling has the character of a public good: it reduces overall societal costs for quantitative energy research as redundant work is avoided, and it improves transparency and reproducibility in electricity system modelling. This paper describes the Open Power System Data platform that aims at realising the efficiency and quality gains of centralised data provision by collecting, checking, processing, aggregating, documenting and publishing data required by most modellers. We conclude that the platform can provide substantial benefits to energy system analysis by raising efficiency of data pre-processing, providing a method for making data pre-processing for energy system modelling traceable, flexible and reproducible and improving the quality of original data published by data providers.
\end{abstract}


\end{frontmatter}


\section{Introduction}
\label{intro}
Applied energy research as well as energy and climate policy advice are often based on quantitative computer models \cite{Pfenninger2014,DeCarolis2017}. Activities in the field of electricity system modelling \cite{Connolly2010,Ringkjob2018} have grown substantially over the last years \cite{Hall2016,Carramolino2017}. This is, among other reasons, driven by increasing shares of wind and solar power in many countries, and the desire to answer various questions of integrating such variable renewable energy sources into power systems and markets. For example, authors of this article have used European electricity models to analyse the impact of renewable energy transitions on various aspects of transmission networks, including congestion management \cite{Kunz2013}, country-specific grid issues \cite{Schlecht2015}, European long-term transmission scenarios \cite{egerer_2016}, merchant interconnectors \cite{GERBAULET2018228} and unscheduled flows under market splitting \cite{Kunz2018}. Other analyses have focused on wholesale prices \citep{Hirth2018}, the changing dispatch of thermal power plants \citep{Schill2017b}, energy storage requirements \citep{Schill2018}, interactions of investments in transmission, storage and thermal plants \citep{Egerer2014}, system effects of electric vehicles \citep{Schill2015}, or optimal configurations of $100\%$ renewable electricity sectors and their respective transition pathways \cite{Wiese2014}.

Since many model-based analyses consider the conditions prevailing in today's power sectors as a starting point, current data is generally needed. While many model applications differ with respect to data requirements --- caused by different model features and different geographical and temporal scopes --- input data on generation capacities as well as time series of load and renewable power generation are relevant input parameters for the majority of electricity system models. Several models also require hourly wholesale prices as input parameters and for some weather data itself is an essential input. The quality of such input data has a major influence on the scientific quality of model-based studies and on their usefulness with respect to policy conclusions.

Yet not only the quality of input data, but also its availability and transparent application are essential for high-quality energy research. In recent years, energy system modelling has been increasingly criticised for its black box character in comparison to other fields \cite{Pfenninger2017}. The openness of code and data are identified as key requirements for energy system models \cite{Pfenninger2018} to comply with scientific standards like improved reproducibility and greater scrutiny \cite{Wiese2018}. By allowing reuse and collaborative development, open models and data can increase productivity and -- through greater transparency \cite{Cao2016} -- also increase credibility in the policy discourse \cite{Dieckhoff2016}. Further, they allow for public participation \cite{Wiese2014}. Since energy system models are highly relevant for real-world policy advice on various aspects of renewable energy transitions, the openness of data and models can also improve public trust and credibility \cite{Pfenninger2017b}. While open model code becomes more common \cite{MORRISON2018} as proven by the growing number of open models \cite{openmod2018}, transparency and openness of data is still not widespread. 

In general, the public availability of European electricity system data has increased significantly during the last years due to rising transparency regulation. An example of this has been the introduction of the ENTSO-E Transparency Platform \cite{ENTSO-E_TP} (see \cite{Hirth2018b} for a review). Yet, data is provided by many different institutions and thus often dispersed over various repositories and formatted in inconsistent ways. Further, data is often poorly documented and sometimes includes obvious errors or missing values. In particular, there are challenges related to:

\begin{itemize}
    \item the identification of appropriate sources
    \item combinations of different data sources
    \item manual downloads of multiple files
    \item merging and harmonising different file formats and inconsistent formatting within files
    \item a lack of a standardised nomenclature and classification of energy sources and technologies
    \item poorly documented original data, in particular a lack of metadata
    \item data quality (e.g inconsistencies, obvious errors, gaps in data)
\end{itemize}

Following \cite{Fowler2017}, we summarise these issues as "friction in the acquisition, sharing, and reuse of research data". Frictionless data, on the other hand, should be easy to find and obtain, be provided in a standardised format and nomenclature and be well documented and error-free.

The quality of such input data has a major influence on the scientific quality of model-based studies and on their usefulness with respect to policy conclusions. Although many models are used to generate insights, not numbers \cite{Huntington1982}, if the numbers that are fed into the model are neither correct nor well documented, no reliable insights can be derived from a model at all. This emphasises the importance for reliable data processing work-flows for the whole energy system analysis.

Furthermore, the collection, processing, maintenance and verification of input data imply substantial work loads for electricity system modellers. So far, such activities are performed by many modelling teams in parallel, and repeated regularly, whenever new data becomes available, i.e.~each year or even more frequently. It would clearly be much more efficient if these tasks would be conducted in a collaborative and/or centralised manner. Likewise, the quality of the processed data is likely to improve by knowledge sharing among electricity system data users.

Thus, the originality of this paper consists in describing a data platform and its underlying methodology, which aims to meet one of the most urgent needs of energy system modellers: transparent, comprehensible and traceable data processing. We provide a holistic method for this purpose, that can be applied by each energy system modeller or community. Furthermore, we demonstrate that the methods presented are actually applicable and well-working to deliver crucial data for applied electricity system research.

Taking into consideration the relevance of input data for electricity system modelling, the need for its transparency and the challenges arising from the current situation, this article provides an overview of the project \textit{Open Power System Data (OPSD)}, which aims at realising the efficiency and quality gains for the modelling community. This is achieved by collecting, checking, processing, aggregating, documenting and publishing data required for European electricity system modelling on an online platform. OPSD was set up in the context of a research project (first phase between 2015-2017) and is in its second phase at the time of writing (2018-ongoing).

The concept and methods of OPSD are described in section \ref{sec:OPSD}, while section \ref{sec:packages} explains the content and value added of the five OPSD data packages. We conclude with a reflection on the current usage of the data platform as well as possible extensions and remaining challenges (Section \ref{sec:discussion}).

\section{Design choices}
\label{sec:OPSD}

To tackle the challenges outlined above, the authors created a central data platform, OPSD \cite{OPSD}, which provides the main input data used for power system modelling at a central location in an aggregated and ready-to-download form. OPSD is a free-of-charge data platform dedicated to electricity system researchers. The data provided on the platform is collected from individual sources, checked, processed, documented and published on the platform.

\subsection{Definition of target group and scope}

The idea behind OPSD is to avoid redundant work when collecting, preparing and aggregating data for energy modelling. Thereby, the project is a service provider to the modelling community. The target group of OPSD are energy professionals such as modellers and analysts in academia, consulting firms and industry. This choice of target group is instrumental in guiding the conceptual decisions for the data platform. We focus on long historical time series rather than high frequency updates of real-time information; and on providing bare CSV files rather than visualisations of data.

The selection of data we provide is based on the data inputs common to most electricity models as outlined in Section \ref{intro}, namely time series data (solar, wind, load and prices), individual as well as aggregated power plant data (conventional and renewable) and weather data. We generally use data from official sources such as from statistical authorities or transmission system operators (TSOs) and only make careful and well-documented changes to the original data.

\subsection{Transparency improves quality and reliability}

To allow for both scientific and policy-relevant use of the provided data, it is crucial to ensure high quality, trustworthiness and reliability. The OPSD concept was designed to meet these criteria. Besides choosing to only rely on data from official sources, the main principle in this regard are open source scripts, transparency of used input data and documentation of both the script as well as the final output.

Along with each of our data packages, we provide the scripts that are used to generate them. These scripts implement all steps including downloading of data from the original sources, processing and reformatting and writing the final output data for download on the OPSD platform. This not only enables data users to read the script to see how the data was treated (such as finding out which method to interpolate missing values was used), but also to take the script and change parts of it in order to implement alternative data processing steps at any point in the pipeline. Users can still benefit from the remainder of the script to generate clean output data.

We put a special emphasis on documentation and chose a two-level approach towards information about data on the platform. On a first level, for users mainly interested in knowing what is to be found in the data files for download, we provide a first short description of data contents along with column-by-column documentation of individual data files directly on the download page. For those who want to dive deeper and know exactly how the data was treated, we use Jupyter notebooks, a format that combines script code and documentation in a single file. The notebooks contain the most detailed documentation of the provided data files and enable users to find out how the data was sourced, treated and packaged up to the final product.

We refrain from making large-scale adjustments to the data in order to remain close to the original sources. This enables data users to choose their individual strategies to treat the data for their use case and trust in the provided data to be close to original sources. However, we do make careful modifications. These include the correction of obvious errors in the original data such as GW-scale rooftop solar plants, alphabetical characters in numerical data fields, missing commas, or aligning of different spellings for the same categorical labels. To remain transparent to the data user about modifications that were applied, a separate marker column marks each instance where data was changed by OPSD.

A frequent problem with data in the power sector is that data that was once available is removed from the original sources, or the location changes. We tackle this problem on the one hand by making a copy of the original input data available on our server next to the data package using it. On the other hand, we try to do better on our side and keep historic versions of data packages online under a stable and clear URL. This is complemented by stable Digital Object Identifiers (DOIs) that we provide to enable referencing individual versions of data packages.

While the majority of data packages on our platform are provided by the core team behind OPSD, we also invite data contributions by third parties. Contributions have to be relevant to our target audience, conform to the technical standards (see Section \ref{subsec:technical}) and be maintained for a prolonged time period to be admitted to the platform.

\subsection{Lightweight and scalable technical architecture}
\label{subsec:technical}

On all technical aspects of our platform, we aim at lightweight decentralised solutions rather than complex centralised structures. We build on established industry standards for open data (Data Packages), existing open-source software solutions (Jupyter Notebooks and Python) and collaborative software development tools (Git and GitHub).

The guiding principle by which we organise the data on our website comes from the Data Package specification \cite{OKI}, a standard developed by Open Knowledge International and the Frictionless Data project \cite{Fowler2017}. The standard is very simple and relies on existing formats. It defines a Tabular Data Package to consist of at least one comma-separated values (CSV) file and a text file containing structured metadata in the JSON format. 

The CSV file format allows for easy and fast download, as no database is required on the server side. Furthermore, since CSV files are human-readable and can be used by almost any software, they are a well-established format. However, some variations of the standards exist, making usage not always intuitive. We tackle this problem by following the Tabular Data Package standard very strictly for our CSV files and by providing Excel and SQLite files in addition.

Metadata is supplied in JSON files, which are both human-readable as well as machine-readable and contain information about properties such as title of the package, sources, contributors and the structure of the CSV files contained in the data package. The machine readability of the metadata files allows users of the data to automatically feed the data into their systems. 

The scripts that we use to automatically download, process, and aggregate data are written in Python. We use Jupyter Notebooks as a single file for coding and documentation. A Jupyter Notebook is a documents that contains live code, equations, visualisations and explanatory text. The notebooks created in this project are available under the open-source MIT license on GitHub \cite{OPSDgithub}.

\subsection{Central vs. decentral data preparation}

Section \ref{intro} has highlighted a number of key issues with data for power system modelling today, often being scattered across multiple sources, non-open and time-consuming to process. Therefore, central data platforms, providing a solution to these problems, are undoubtedly beneficial in regards to cutting the time and cost involved in data preparation. However, centralisation of data preparation also entails certain risks, which we outline and discuss below, as they are crucial to the key choices made in creating the platform.

First, a major risk in central data provision concerns error propagation. A single error in data preparation on the central data platform could propagate into many studies based on the centrally provided data. We tackle this risk by providing our scripts open source, so that others are invited to use and scrutinise the scripts (which, according to the detailed feedback we get, actually takes place). This makes error-detection easier than on closed-source platforms, where users do not get to see the programming code. Furthermore, we mark all instances where we intentionally modify and correct data, which points users at the most critical data points to scrutinise.

Second, there might be a bias in data selection. While for some data sets many competing sources are available (such as installed power plant capacity figures being available from statistical offices, energy ministries and TSOs), the data source selected by a central platform such as OPSD might be over-represented in final data usage by scientists and analysts. We tackle this issue by providing, where available, competing data sources so that users can compare them against each other. This is especially the case for the national generation capacity data package (see Section \ref{subsec:national generation capacity}), but also concerns the time series data package (see Section \ref{subsec:time series}) regarding load values from two different ENTSO-E sources.

Third, a central data platform might lead to a reduction of the plurality of data preparation methods in use before. By centralising model data provision, all users of the centrally provided data will base their studies on the selected approaches instead of implementing their own methodologies in regard to interpolation of missing values or detection of improbable data outliers, which might lead to a bias if there are issues with the chosen treatment. We address this issue by applying data modifications carefully only and in cases where a correction is not possible and mark improbable data items, leaving it to the user to do final modifications.

In summary, we argue that while there are risks from centralising data preparation, these are to a large degree addressed by the design choices we take.

\subsection{Licensing}
Ideally, the data available on OPSD would be provided under an open licence such as those from the creative commons family. However, a prerequisite for granting an open license would be that the providers of original data sources gave their permission or were themselves using open licenses, which is hardly the case (exceptions are Danish and French TSOs Energinet.dk \cite{Energinet_DataService} and RTE \cite{RTE_ODRE}). We have sent enquiries to data providers asking to republish their data under the Creative Commons Attribution 4.0 International license. While some providers granted their permission, others refrained or did not reply. Singling out and licensing the data from providers who gave permission would be possible, but highly impractical, which is why we decided against it.

\subsection{Maintenance and longevity}
All too often with the end of funding for an academic project also the outputs it provides fade. Therefore, we have tackled the issue of long-term maintenance and future updates in our design choices already.

Three design choices we made are instrumental in promoting longevity. Firstly, by doing our data preparation script-based and providing those scripts as open source, we make it very easy for others to continue on from the state we might ever leave the project in, in the unlikely case that we do not manage to secure a continuation beyond 2020. The choice of an open-source license also makes transitioning of the whole OPSD project to other institutions possible, as no intellectual property rights stand in the way. Secondly, by choosing a lightweight and decentralised IT approach that requires only a very basic and cheap shared-hosting environment, we keep the hosting costs below the price of a cup of coffee per month, enabling us to make long-term commitments to keep hosting alive. Thirdly, by choosing to provide Digital Object Identifiers (DOIs) for our data packages, we were obliged to sign a contract with the DOI issuer and commit to a long-term availability of the provided data. This also helps us to signal the stability and longevity we strive for to the users of our platform.

\section{Five data packages}
\label{sec:packages}
\subsection{Overview}
For the illustration of our work on and methods for tackling the challenges of input data for electricity system modelling, we describe scope, coverage and value added for each of the five currently available OPSD data packages. These cover a large share of input data required by electricity system models:
\begin{itemize}
    \item Conventional power plants -- Lists of conventional power plants and units
    \item Renewable power plants -- Lists of renewable energy power stations
    \item National generation capacity -- An overview of generation capacities by technology and country from different original sources
    \item Time series -- Load, wind, solar, and price data in hourly resolution
    \item Weather data -- Hourly geographically aggregated weather data for Europe
\end{itemize}

\begin{figure}[!h]
    \centering
    \includegraphics[width=1\textwidth]{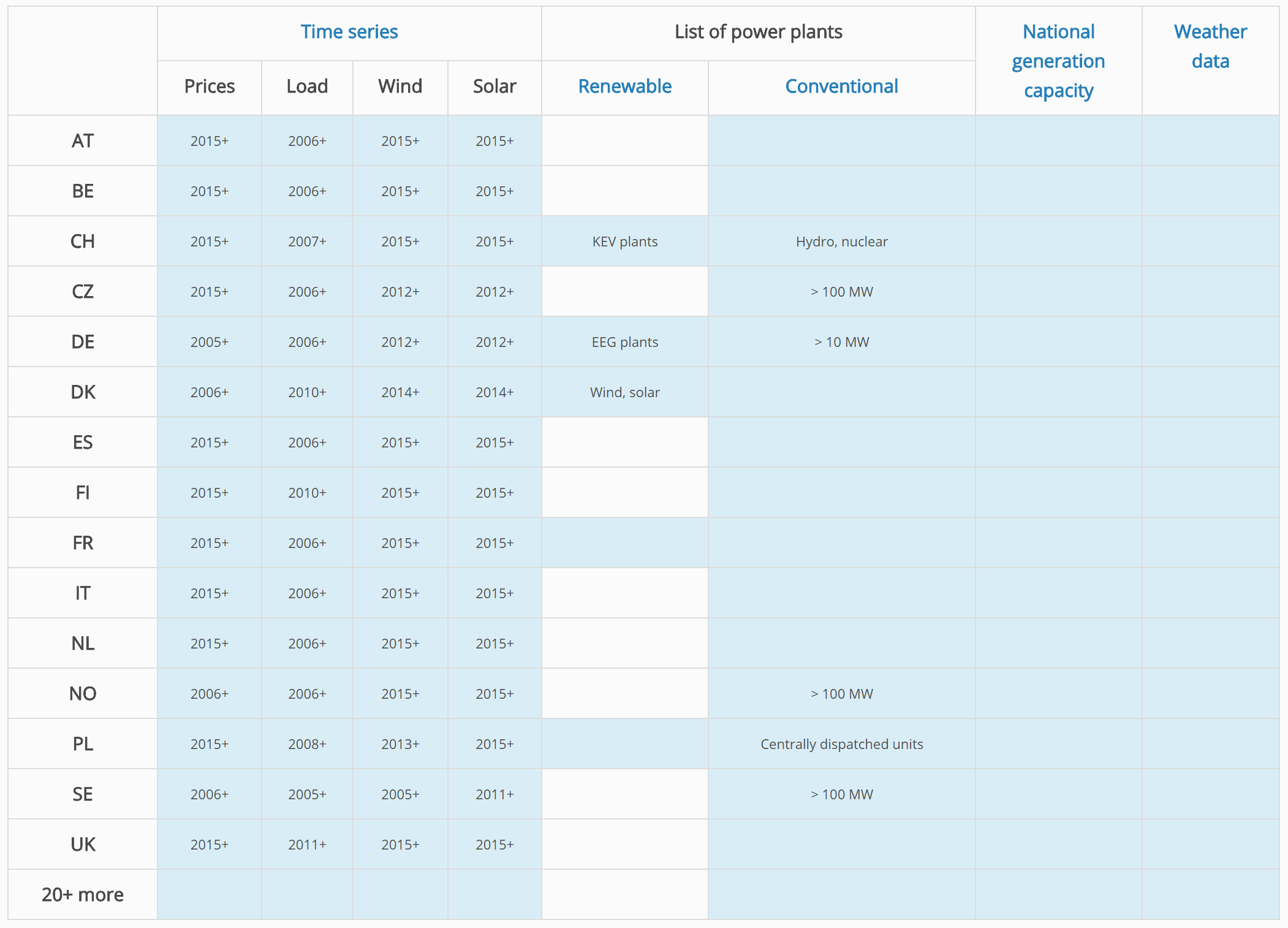}
    \caption{Overview of the current data availability (as of May 2018).}
    \label{fig:overview}
\end{figure}

Due to a varying availability of original data, the geographical as well as the temporal coverage of the resulting data packages differ. Therefore, an overview of coverage is provided in the data availability matrix (Figure \ref{fig:overview}). Countries covered in the different data packages are indicated by blue coloured cells. The cell content further specifies the earliest year of availability (time series) or the type of power plants covered (renewable and conventional).

\begin{figure}[!h]
    \centering
    \includegraphics[width=1\textwidth]{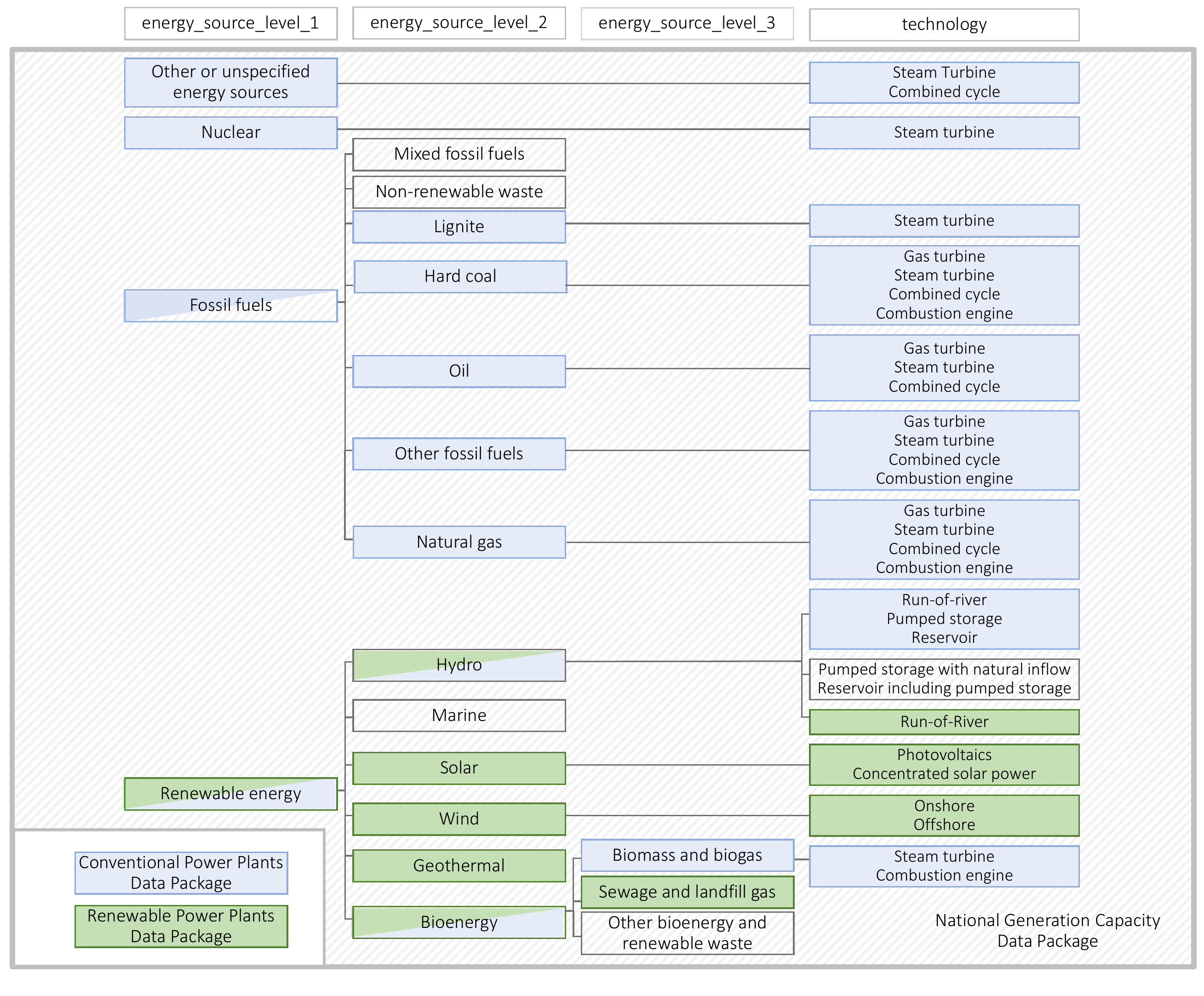}
    \caption{Classification of energy sources and technologies}
    \label{fig:energy_structure}
\end{figure}

A time-consuming barrier for input data handling from different sources are divergent naming conventions and classifications from different original data sources. Thus, an essential contribution of OPSD for smoother input data handling and better comparability of original data is a consistent classification of energy sources and technologies (see Figure \ref{fig:energy_structure}) which we consistently apply to the data packages where this is relevant, i.e. the first three in the list above. On the first level, we generally distinguish between renewable energy sources, fossil fuels, nuclear and other or unspecified sources. The second level contains different renewable sources such as wind and solar power as well as different fossil fuels. The third level is necessary to distinguish between different forms of bio energy.  

In the following, we describe the content, scope and particular features of each data package. Please note that only the most relevant of the large amount of original data sources are mentioned while a full list can be found on the OPSD platform itself \cite{OPSD}.

\subsection{Conventional power plants}
\paragraph{Content and Scope}
This package contains data on conventional power plants for Germany and selected other European countries. It covers nuclear reactors and thermal plants fuelled by lignite, hard coal, natural gas or oil, as well as hydro power and pumped hydro storage. The package focuses on individual power plants, or power plant blocks, and their technical characteristics. These include installed capacity, main energy source and type of technology. Where available, we also include capabilities with respect to combined heat and power (CHP) generation, commissioning and closure years as well as geographical information and Energy Identification Codes (EIC). We also provide information on thermal efficiency of individual plants, either by means of desk research or based on an estimation method described in \cite{Egerer.2014}.

The main data source for Germany is a detailed power plant list provided by the federal network regulator ({BNetzA} \cite{BNetzA}). It is complemented with a list provided by the German Environment Agency (UBA \cite{UBA}) which contains additional information on CHP, installed gross capacities, and turbine types. Sources for other countries include detailed power plant lists provided by TSOs, ministries, or associations and market participants. For countries without system-wide power plant lists the lists have been manually combined from multiple official sources.

While we implemented a single data structure for all countries covered, the package consists of two separate Jupyter Notebooks and two separate output files: one for Germany and one for the other countries. This is because the original German data provides many more details compared to the other countries, and also requires a specific merging activity of the lists provided by {BNetzA} and UBA.

\paragraph{Geographic coverage}
Geographically, the data package focuses on Germany and its neighbouring countries. The data package currently covers Austria, Belgium, the Czech Republic, Denmark, Germany, Finland, France, Italy, the Netherlands, Norway, Poland, Spain, Sweden, Switzerland, Slovakia, Slovenia, and the United Kingdom. Due to varying availability and quality of original data, not all information is available for each country.

\paragraph{Value added}
This data package is designed to benefit electricity system modellers in several ways. First, we provide a single data structure for many countries, including a harmonisation of fuels and technologies, the latter sometimes inferred from other information provided in the original data or based on own assessments. Specifically for Germany, merging the two official lists provided by {BNetzA} and UBA adds value with respect to CHP information, net / gross capacities, as well as commissioning and shut-down years. Further, complementary information which is not included in the primary sources but  researched manually, i.e.,~geo coordinates, EIC codes, and thermal efficiency, should be of use for many modellers.

\subsection{Renewable power plants}
\paragraph{Content and Scope}
This data package contains a list of existing solar, wind, bio, run-of-river, geothermal and other renewable power plants. Due to different data and parameter availability in the countries, one list is provided per country. These are of different accuracy and partly include different parameters. However, all lists state energy source and technology, electrical capacity and data source. If available in the original data, the respective system operators (transmission and distribution), support scheme ID or technology-specific parameters like hub height for wind turbines are included. While a high amount of parameters and units (more than 1.8 million power plant entries) are available for Germany, these only cover plants eligible under the renewable support scheme (EEG). Also for Switzerland just the supported plants are provided (Swiss feed-in tariff KEV) since these are well documented and information is publicly provided. For Denmark, detailed data on solar and wind power plants including rotor diameter etc. is provided, but no other technologies. Aggregated capacity per energy source and municipality is stated in the lists for Poland and France.

Most original data sources either provide conventional or renewable power plants. However, especially with respect to hydro power, there is an overlap between sources, and harmonisation is challenging. In Germany, in the last decades, several data providers (four TSOs and the federal agency BNetzA) report differently on hydro power plants depending on various laws, regulation and publication requirements. Combining different sources partly leads to double-counting of hydro power plants in renewable and conventional power plant lists, due to inconsistent statistic counting methods. With the objective to provide data packages that can be directly applied by the modellers, we eliminated the overlap to some extent by clarifying which energy sources and which technologies are covered by the conventional and renewable data package as illustrated in Figure \ref{fig:energy_structure}.

\paragraph{Geographic coverage} Currently, lists of German, Danish, Swiss, Polish and French renewable power plants are provided. Since the location of renewable power plants is important for resulting feed-in of wind and solar, the geographic location is provided wherever information was available. This is the case for Germany, Denmark and France. These coordinates are of different accuracy depending on the original source being derived from coordinates in UTM-format (BNetzA), zip-codes (German TSOs) or districts.

\paragraph{Value added}
The amount of renewable energy units by far exceed the conventional ones due to smaller unit size, which makes processing tedious and manual verification on a unit-by-unit base impossible. Furthermore, naming conventions and classification of renewable energy technologies and sources differ significantly between data providers. The added value of the script is thus bringing it to the same energy source structure as well as to the same formatting. Although some parameters differ between the country scripts, the country lists can be consistently combined for a subset of essential parameters. Furthermore, inconsistent data points (like e.g. rooftop solar installations of several GW) are marked and can thus easily be filtered out.

An additional output of this data package and value added are historic daily time series of the installed renewable capacities per energy source type for Germany. These are applied in the time series script: Since solar and wind feed-in time series are derived from real feed-in data, feed-in levels tend to increase over the year due to increasing installed capacity. For deriving relative, normalised hourly profiles, the corresponding installed capacity of the respective day is required. The creation of this output requires the date of installation from the original data source and is thus only provided for those countries for which this information is available. Until November 2018, this is only the case for Germany.

\subsection{National generation capacity}
\label{subsec:national generation capacity}
\paragraph{Content and Scope}
This data package complements the \textit{Conventional power plants} and \textit{Renewable power plants} packages by providing aggregated numbers on installed capacities for all types of generators. To create this package, we compiled a broad range of sources and structured them according to the harmonised classification of energy sources and technologies provided in Figure \ref{fig:energy_structure}. This proved to be challenging because of very heterogeneous reporting conventions across national statistics. Overall, we provide more than $1500$ annual country statistics from four international and $25$ national sources, i.e.~around four statistics for each country covering several years each. As for the international sources, we use data provided by ENTSO-E and EUROSTAT, the latter going back to 1990. National data sources include statistical offices, ministries, TSOs and industry associations. While the international sources are automatically processed, the national sources generally require manual compilation.

\paragraph{Geographic coverage}
The package provides data on 39 countries and thus virtually comprises all European countries. 25 of these countries are not only covered by international sources, but also complemented by individual country statistics. For example, 2015 data for France comprises four distinct data sources. 

\paragraph{Value added}
The main output of this package is a single data file that includes installed generation capacities for all European countries, with a harmonised coverage of energy sources and generation technologies. This is valuable for all electricity system models that do not require much technological or geographic detail, for example linear models that consider whole countries as nodes. Additional value is created by contrasting different sources for a given country and year, located in adjacent columns of the data file. This enables the user, inter alia, to draw conclusions on the quality and coverage of specific sources or to infer missing values. For example, overall generation capacity installed in France in 2015 varies between $105$GW and $129$GW across the different sources, with particularly large variation among renewable sources.

To name another benefit, the temporal and geographical coverage of this package is much better compared to the other two data packages covering conventional and renewable generators. Not least, we also provide a strict distinction between values that are actually zero and such that are not available or not specified in national / international statistics.

\subsection{Time series}
\label{subsec:time series}

\paragraph{Content and Scope}
The \textit{Time series} data package contains highly granular time series data for a number of parameters. Included are power consumption (load), day-ahead power prices, as well as three types of time series related to generation by variable renewables (wind and solar). The wind and solar time series include historical forecast generation, actual generation and installed production capacity. The latter is calculated from the \textit{Renewable power plants} package and is required to calculate renewable generation profiles as share of installed production capacity.

Depending on the national market set-up, the time series have a resolution of either 15 minutes (i.e. Austria, Germany and some neighbours), 30 minutes (i.e. the British Isles) or 60 minutes (all other countries).

While from 2015 onward, many of the data are available from the ENTSO-E Transparency Platform, earlier records have to be obtained through individual national sources. Data are extracted mostly from national TSO's websites, ENTSO-E's "Monthly statistics data collection" as well as ENTSO-E Transparency. 

Many original data sources report data by referring to the local time, in most cases Central European (Summer) Time (CET/CEST). Transitions between summer daylight saving time and winter time are a source of confusion, since they imply jumping an hour in March and reporting an hour twice in October. We thus consistently convert all time stamps to Coordinated Universal Time (UTC), which avoids daylight saving time.

Since power system modelling usually requires time series data to be complete for at least an entire year, data gaps of up to two hours are filled by applying linear interpolation. In these cases, tables are annotated with a marker, allowing users to trace back original data manipulation, while longer periods of missing data are left as-is.

\paragraph{Geographic coverage}
Overall, 35 countries are covered while varying between parameters and improving in more recent years. While load data is published for all European countries, availability of renewable generation data varies between countries. Historic records of day-ahead prices are freely available since 2015, when the ENTSO-E Transparency Platform was introduced.

\paragraph{Value added}
This package relieves researchers of the burden to collect time series data from many different sources, each using different formats and ways to organise data: Some TSOs allow downloading one file containing all available data (Amprion), while others require downloading and combining monthly (TenneT) or even daily files (PSE). The time series data package combines all data with the same temporal resolution in one file each, facilitating its use for modellers. Additionally, 15 and 30 minutes data are aggregated to hourly time series since this is the resolution required by most power system models and enables one file with maximal geographic coverage while keeping the file size manageable.

\subsection{Weather data}

\paragraph{Content and Scope}
The \textit{Weather data} package provides a script to automatically access, subset, download and process wind, solar, temperature and air data based on the MERRA-2 dataset \cite{Merra2,Rienecker2011} made available by NASA. This data package differs significantly from the other data packages as the size of the data to be downloaded can reach GB and TB if a large amount of parameters, a long time span and a huge geographical area are chosen. Thus, instead of offering the download of ready-made data packages, the strategy here is to provide a documented methodological Python script that can be run on the user's computer. This script serves as a documented tutorial and enables the user to specify the geographical area and the time span for which the data shall be downloaded and processed. Only a small sample data set (Germany, 2016) is available for direct download.

Besides connection to the database, download and conversion of the three-dimensional NetCDF4-format to CSV and SQL-output, the script includes the calculation of the input parameters relevant for electricity system modellers. For example, the wind speed is calculated from the north and east pointing wind vectors provided in the original MERRA-2 data set. Resulting parameters covered by the script are:
\begin{itemize}
 \item wind velocity in 2, 10 and 50 meters height as well as roughness length
 \item solar radiation
 \item temperature
 \item air density and pressure
\end{itemize}

\paragraph{Geographic coverage}
The MERRA-2 data set offers reanalysis data with a worldwide coverage in a resolution of 0.625$^{\circ}$ x 0.5$^{\circ}$ which correlates to approximately 50 x 50 km in Central European latitudes. By specifying the northeastern and southwestern corner coordinates in the script, a rectangular area can be chosen.

\paragraph{Value added}
Due to the rising influence of wind and solar feed-in for the power system, their realistic representation and thus weather parameters are crucial for electricity system models. The OPSD weather data script fills a gap between original meteorological data and ready-made feed-in time series. On the one end of the spectrum, the direct usage of original data sets like MERRA-2 with hundreds of different weather parameters to choose from as well as a tedious manual download and processing requires specific knowledge to retrieve the relevant data for electricity system modelling. Additionally, the NetCDF-file format preferred by meteorologist due to its multidimensional structure is hard to handle and not directly compatible with data formats preferred by energy system modellers. On the other end, ready-made feed-in time series as provided in the data package described in Section \ref{subsec:time series} are less flexible for the automatic adaption to own modelling code and are restricted by fixed assumptions made for the calculation of the resulting electricity feed-in, e.g. the hub height or type of wind turbines. Using CSV instead of original NetCDF-files implies the downside of less efficiency in terms of organising and storing the data. But since this is a major limitation for electricity system modellers who require decades of historical weather data, and who carry out own feed-in calculations, the OPSD script lowers the barrier of utilising the wealth of available reanalysis weather data by making them more accessible in CSV-format.
        
\section{Discussion and Outlook}
\label{sec:discussion}
The OPSD platform has been designed to provide a service to the electricity system modelling community. It provides a wide range of data at one place, is easily accessible, clean and ready-to-use, permanently available and version-controlled. The large number of users --- around $100,000$ unique visitors during 2017) --- and, more importantly, the amount of research that makes use of OPSD 
--- 26 published papers by the time of writing since the go-live in late 2016, out of which 12 are published papers in high quality journals indexed in the SCI/SSCI 
\cite{Klein2017,ElAmary2018,Gonzales2018,Zaidi2018,Gotzens2018,Schill2017,Schill2017b,Zerrahn2018,Olauson2018,Brown2018b,Gambardella2018,Portela2017} 
and 14 in other journals, conference papers, books and grey literature \cite{Masum2018, Mueller2017,Kendziorski2017,Fusco2017,Tafarte2017,Lee2018,Riepin2018,Nacken2017,Robinius2018,Osegi2018,Sharma2018,Amme2018,Mueller2018,Yaslan2017} ---
suggest that the platform fulfils a need and provides value to electricity system modellers. 

The main contribution of this paper is to provide a systematic overview of the OPSD platform and its main methodological approaches. This platform addresses a big challenge for energy system modellers: to process and document massive amounts of input data from various sources in a flexible, but still reproducible way. It thus helps to overcome one of the main barriers to improving transparency in modelling, which researchers have urgently called for \cite{Pfenninger2018}. 

Clearly, there is room for improvement. Planned future work includes expanding the geographic scope of the platform, adding additional data items, and improving usability.

An ongoing challenge for projects like OPSD is the academic incentive structure. Collecting, cleaning, aggregating and providing data is essential, everyone agrees. However, it is challenging to attract funding for these kinds of projects and even more difficult to publish journal articles on data processing and aggregation. As a consequence, open data, like many other public goods, tends to remain short in supply.

In the long term, we hope that OPSD provides additional benefits by serving as a proof of concept. It may help to convince the providers of data --– system operators, statistical offices, generation companies, and public authorities --– to provide data that is complete and can be easily used for research. With machine-readable data and metadata that is packaged and version-controlled, along with quality and consistency checks and detailed data documentation, the electricity sector would make a huge step towards frictionless data.

\section*{Acknowledgements}
The OPSD project has been supported by the German Federal Ministry for Economic Affairs and Energy (BMWi) in its first phase 2015-2017 under grant number 03MAP324. Since 2018, it is maintained and amended as part of the project "Open Source Energiewende" (060/17) for BMWi. The authors also thank all colleagues contributing to the platform, the open energy modelling initiative and all users of OPSD providing feedback. 
\section*{References}

\bibliography{mybibfile}

\end{document}